\documentstyle[prl,aps,multicol,epsf]{revtex}
\begin {document}
\draft
\title{ Heating of a two-dimensional electron gas
by the electric field of a surface acoustic wave}
\author{ I. L. Drichko, A. M. D'yakonov, V. D.
Kagan, A. M. Kreshchuk, T. A. Polyanskaya, I. G.
Savel'ev, I. Yu. Smirnov, and A. V. Suslov}
\address{ A. F. Ioffe Physicotechnical Institute.
Russian Academy of Sciences, 194021
St.Petersburg, Russia}

\date{\today}
\input{psfig}
\maketitle

\begin{abstract}

The heating of a two-dimensional electron gas by
an rf electric field generated by a surface
acoustic wave, which can be described by an
electron temperature $T_e$, has been
investigated.
It is shown that the energy balance of the
electron gas is determined by electron scattering
by the piezoelectric potential of the acoustic
phonons with $T_e$ determined from measurements
at
frequencies $f$= 30 and 150 MHz. The experimental
curves of the energy loss $Q$ versus $T_e$ at
different SAW frequencies depend on the value of
$\omega \bar{\tau}_{\epsilon}$, compared to 1,
where $ \bar {\tau}_{\epsilon}$ is the
relaxation
time of the average electron energy. Theoretical
calculations of the heating of a two-dimensional
electron gas by the electric field of the surface
acoustic wave are presented for the case of
thermal electrons ($\Delta T \ll T$). The
calculations show that for the same energy losses
$Q$ the degree of heating of the two-dimensional
electrons (i.e., the ratio $T_e/T$) for
$\omega \bar{\tau}_{\epsilon}>1$ ($f$= 150 MHz) is less
than for $\omega \bar{\tau}_{\epsilon}<1$ ($f$=30
MHz). Experimental results confirming this
calculation are presented.

\end{abstract}
\bigskip
\pacs{PACS numbers: 72.50.+b; 73.40.Kp}

\begin{multicols}{2}
\section{INTRODUCTION}

The investigation of nonlinear (with respect to
the input power) effects in the absorption of
piezoelectrically active ultrasonic waves,
arising
due to the interaction of the waves with three-
dimensional electron gas (in the case of
Boltzmann
statistics), has shown that the mechanisms of the
nonlinearity depend on the state of the
electrons.
If electrons are free (delocalized), then the
nonlinearity mechanism for moderately high sound
intensities is usually due to the heating of the
electrons in the electric field of an ultrasonic
wave. The character of the heating depends on the
quantity $\omega \tau_{\epsilon}$, where $\omega$
is the sound frequency, and $\tau_{\epsilon}$ is
the energy relaxation time \onlinecite{1,2}. If
the electrons are localized, then the
nonlinearity
mechanism is due to the character of the
localization (on an individual impurity or in the
wells of a fluctuation potential). In \cite{3},
it
was shown that in the case where the electrons
are
localized on individual impurities, the
nonlinearity
was determined by impurity breakdown in the
electric field of the sound wave. When the
electrons occupied the conduction band as a
result
of this effect, their temperature started to grow
as a result of heating in the electric field of
the wave \cite{4}.

The study of structures with a two-dimensional
electron gas (2DEG) opens up a unique possibility
of studying in one series of measurements
performed on the same sample the mechanisms of
nonlinearity in delocalized and localized
electron
states, since under quantum Hall effect
conditions
both states are realized by varying the magnetic
field. The change in the absorption coefficient
for a piezoelectrically active surface acoustic
wave (SAW) interactimg with a 2DEG as a function
of the SAW intensity in GaAs/AlGaAs structures
was
previously observed in \cite{5} and \cite{6} only
in the magnetic field range corresponding to
small
integer filling numbers-the quantum Hall effect
regime, when the two-dimensional electrons are
localized. The authors explained the data which
they obtained by heating of a 2DEG.

In the present paper we report some of our
investigations concerning nonlinear effects
accompanying the interaction of delocalized two-
dimensional electrons with the electric field of
a
SAW for the purpose of investigating nonlinearity
mechanisms.

\section{EXPERIMENTAL PROCEDURE}

We investigated the absorption coefficient for a
30-210 MHz SAW in a two-dimensional electron gas
in $GaAs/Al_{0.75}Ga_{0.25}As$ heterostructures
as
a function of the temperature in the range
$T\simeq 1.4-4.2K$ in the linear regime (the input
power did not exceed $10^{-7}$W) and the SAW power
at $T=1.5K$ in magnetic fields up to 30 kOe.
Samples studied    previously    in    \cite{7}
with    Hall density  $n_s^H=6.7\cdot10^{11}cm^{-2}$
and mobility  $\mu_H=1.28 \cdot 10^5 cm^2/(Vs)$ at T=4.2 K were used for the
investigations. The technology used to fabricate
the heterostructures is described in \cite{8}
and
the procedure for performing the sound absorption
experiment is described in \cite{7}. Here we only
note that the experimental structure with 2DEG
was
located on the surface of the piezodielectric
(lithium niobate $LiNbO_3$), along which the SAW
propagates. The SAW was excited in a pulsed
regime
by sending radio pulses with filling frequency
30-
210 MHz from an rf oscillator into the excited
interdigital transducer. The pulse duration was
of
the order of 1$\mu s$ is and the pulse repetition
frequency was equal to 50 Hz. In the present
paper
the SAW power is the power in a pulse.

An ac electric field with the frequency of the
SAW, which accompanies the deformation wave,
penetrates into a channel containing two-
dimensional electrons, giving rise to electrical
currents and, correspondingly, Joule losses. As a
result of this interaction, energy is absorbed
from the wave. The SAW absorption in a magnetic
field is measured in the experiment. Since the
measured absorption is determined by the
conductivity of the 2DEG, quantization of the
electronic spectrum, which leads to Shubnikov-de
Haas oscillations, gives rise to oscillations in
the SAW potential as well.

\section{EXPERIMENTAL RESULTS AND ANALYSIS}

Curves of the absorption coefficient $\Gamma$
versus the magnetic field $H$ are presented in
Fig.1 for different temperatures and powers of
the
30-MHz SAW. Similar curves were also obtained for
other SAW frequencies. The character of the
curves
$\Gamma (H)$ is analyzed
in \cite{7}. The absorption maxima $\Gamma_{max}$
as a  function of the magnetic field for
$H<25$kOe
are equally spaced as a function of $1/H$, and
the
splitting of the maxima $\Gamma(H)$ for $H>25$ kOe
into two peaks with the values of $\Gamma_M$ at
the maxima \footnote{In \cite{7} it is shown that
the values of $\Gamma_M$ do not depend on the
conductivity of the 2DEG, and that they are
determined, within the limits of the experimental
error, only by the SAW characteristics and the
gap
between the sample and $LiNbO_3$}
is due to the relaxational character of the
absorption. The temperature and SAW power
dependences of $\Gamma$, shown in Figs.
2 and 3, were extracted from the experimental
curves of the same type as in Fig.1 for the
corresponding frequencies in a magnetic field
$H<25$kOe for large filling numbers
$\nu=n_shc/2eH>7$.

Figure 2 shows the temperature dependence of the
quantity $\Delta\Gamma=\Gamma_{max}-\Gamma_{min}$
measured in the linear regime at a frequency of
150 MHz in different magnetic fields. Here
$\Gamma_{max}$ and $\Gamma_{min}$ are the values
of $\Gamma$ on the upper and lower lines, which
envelop the oscillatory dependence $\Gamma(H)$
for
$H<25$kOe. Figure 3 shows $\Delta\Gamma$ versus
$P$-the power of the SAW (frequency 150 MHz) at
the oscillator output at $T$=1.5K. We see from
Figs. 2 and 3 that $\Delta\Gamma$ decreases with
increasing temperature and with increasing SAW
power.

In \cite{7} it was shown that in the range of
magnetic fields where the quantum Hall effect is
still not observed (in our case $H<25$ kOe) the
dissipative conductivities are

$\sigma_{XX}^{ac}=\sigma_{XX}^{dc}$,

where $\sigma_{XX}^{dc}$ is the conductivity
calculated from the measured dc resistivities
$\rho_{xx}(H)$ and $\rho_{xy}(H)$, and
$\sigma_{XX}^{ac}$ is the conductivity found from
the absorption coefficient $\Gamma(H)$ measured
in
the linear regime. This result gave us the basis
for assuming that in this range of magnetic
fields
the electrons are in a delocalized state. As we
have already indicated in the introduction, we
shall analyze here nonlinearity only in this
case.

In a previous work \cite{9} we showed that if the
electrons are delocalized, then the
characteristics of the 2DEG, such as the carrier
density $n_s$, the transport relaxation time
$\tau$, and the quantum \footnote{We take this
term to mean the so called escape time $\tau_0$
which is inversely proportional to the almost
total scattering cross section \cite{10}. In
experiments on quantum oscillations it is defined
as time $\tau_0=\hbar/2\pi T^*$, where $T^*$ is the
Dingle temperature.}
relaxation time, can be determined from the
magnetic field dependences $\sigma_{XX}^{ac}(H)$.
In addition, the mobility $\mu=e\tau/m$ and the
concentration $n_s$ at $H=0$ are close to the
values obtained from dc measurements: the Hall
density and mobility of the electrons, as well as
$n_s$ found from the Shubnikov-de Haas
oscillations. For this reason, it was natural to
assume that $\Gamma$ depends on the SAW power,
just as in the static case, because of the
heating
of the 2DEG but in the electric field of the SAW.
The heating of the 2DEG in a static electric
field
in similar heterostructures was investigated in
\cite{11,12,13,14,15}. In those papers it was shown that
at
liquid-helium temperatures the electron energy
relaxation processes are determined in a wide
range of 2DEG densities by the piezoacoustic
electron-phonon interaction under small-angle
scattering and weak screening conditions.

We shall employ, by analogy with \cite{11,12,13,14},
the
concept of the temperature $T_e$ of two-
dimensional electrons and determine it by
comparing the curves of the absorption
coefficient
$\Gamma$ versus the SAW power with the curves of
$\Gamma$ versus the lattice temperature $T$. Such
a comparison makes it possible to establish a
correspondence between the temperature of the
two-
dimensional electrons and the output power of the
oscillator. The values of $T_e$ were extracted by
two methods: 1) by comparing the curves of the
amplitude of the oscillations
$\Delta\Gamma=\Gamma_{max}-\Gamma_{min}$ versus
the temperature $T$ (Fig. 2) and versus the power
$P$ (Fig. 3) for the same value of the magnetic
field $H$; 2) by comparing curves of the ratios
$\Gamma_{max}/\Gamma_M=f(T)$ and
$\Gamma_{max}/\Gamma_M=f(P)$ versus the lattice
temperature $T$ and the power $P$. Here the
values
of $\Gamma_{max}(T)$ and $\Gamma_{max}(P)$  were
also taken for the same value of $H$, and
$\Gamma_M$ is the absorption at $H$=28 kOe (Fig.
1). The use of the ratios instead of the absolute
values of $\Gamma$ decreased the effect of the
experimental variance in $\Gamma$  on the error
in
determining $T_e$. As a result, the accuracy in
determining $T_e$ by these two methods was no
worse than $10\%$.

To determine the absolute energy losses as a
result of absorption of SAW in the case of
interaction with electrons ($\bar{Q}$), the
following calculations must be performed. The
intensity $E$ of the electric field, in which the
two-dimensional electrons of the heterostructure
are located during the propagation of a SAW in a
piezoelectric material placed at a distance $a$
from
a high-conductivity channel, is

\begin{eqnarray}
\  |E|^2=K^2\frac{32\pi}{\nu}(\epsilon_1+\epsilon_0)
\frac{bqexp(-2qa)}
{1+[(4\pi \sigma_{xx}/\epsilon_s v)c]^2}W,\label{eq:1}
\end{eqnarray}

where $K^2$ is the electromechanical coupling
constant; $\nu=3.5 \cdot 10^5$ cm/s and $q$ are,
respectively, the velocity and wave number of
sound in $LiNbO_3$; $a$ is the width of the
vacuum
gap between the sample and the $LiNbO_3$ plate;
$\epsilon_0$, $\epsilon_1$ and $\epsilon_s$ are
the permittivities of free space, $LiNbO_3$, and
the semiconductor with the 2DEG, respectively;
and
$W$ is the input SAW power scaled to the width of
the sound track. The functions $b$ and $c$ are

\begin{eqnarray}
b=( \epsilon_1^+ \epsilon_s^+- \epsilon_1^- \epsilon_s^- \exp(-2qa))^{-2},
\nonumber \\
c=\frac{1}{2}
(1+b^{1/2}[\epsilon_1^+\epsilon_s^- -\epsilon_1^-\epsilon_s^+\exp(-2qa)]),
\nonumber \\
\epsilon_1^+=\epsilon_1+\epsilon_0,
\epsilon_s^+=\epsilon_s+\epsilon_0,\nonumber \\
\epsilon_1^-=\epsilon_1-\epsilon_0, \epsilon_s^-
=\epsilon_s -\epsilon_0.
\nonumber
\end{eqnarray}
The magnitude of the electric losses is defined
as
$\bar{Q}=\sigma_{xx}E^2$. Multiplying both sides
of Eq.(1) by $\sigma_{xx}$, we obtain
$Q=4W\Gamma/n$, where $\Gamma$ is the absorption
measured in the experiment. The power $W$ at the
entrance to the sample is not measured very
accurately in acoustic measurements. The problem
is that this quantity is determined by, first,
the
quality of the interdigital transducers; second,
by the losses associated with the mismatch of the
line that feeds electric power into the
transmitting transducer as well as the line that
re moves electrical power from the detecting
transducer, where the losses in the receiving and
transmitting parts of the line may not be the
same; and, third, by absorption of the SAW in the
substrate, whose absolute magnitude is difficult
to
measure in our experiment. The effect of these
losses de creases with frequency, so that in
determining $W$ at 30 MHz we assumed that both the
conversion losses for the transmitting and
receiving transducers as well as the losses in th
transmitting and receiving lines are identical.
The total losses were found to be $\Delta P$=16
dB, if SAW absorption in the heterostructure
substrate is ignored. If it is assumed that nonlinear
effects at 150 MHz start at the same value
of $\bar{Q}$ as a 30 MHz, then the "threshold" value of
$\bar{Q}$ at which the deviation of
$\Gamma_{max}(\bar{Q})/\Gamma_M$ at 30 MHz from a
constant value be comes appreciable [we recall
that $\Gamma(H)/\Gamma_M \sim 1/\sigma_{xx}(H)$
in
the region of delocalized electronic states,
i.e.,
$H<$25 kOe \cite{7}] can be used to determine the
total losses at 150 MHz. An estimate of the total
losses by this method at 150 MHz give; $\Delta P$=18 dB.
Therefore, the power $W$ at the
entrance
to the sample is determined by the output power
$P$ of the oscillator taking into account the
total losses $\Delta P$.

With the results of \cite{11,12,13,14} in mind, we
constructed the curves

\begin{eqnarray}
Q=\bar{Q}/n_s=f(T_e^3-T^3),
\nonumber
\end{eqnarray}

which correspond to the energy balance equation
in
the cast of the interaction of electrons with the
piezoelectric potential of the acoustic phonons
(PA scattering) under the condition of weak
screening at frequencies of 30 and 150 MHz in
different magnetic fields

\begin{eqnarray}
Q_{PA}= e\mu E^2=A_3 (T_e^3-T^3).
\label{eq:2}
\end{eqnarray}

But since the condition for weak screening was
not
satisfied for this sample, the curves $Q=f(T_e^5-
T_0^5)$, corresponding to the energy balance
equation in the case of PA scattering but with
the
condition of strong screening for the same
frequencies 30 and 150
MHz and the same magnetic fields, were also
constructed:

\begin{eqnarray}
Q_{PA}=A_5(T_e^5-T_0^5),
\label{eq:3}
\end{eqnarray}

A least-squares analysis showed that the
expression (3) gives a better description of the
experimental curves. Figure 4 shows the
experimental points and theoretical curves of the
expressions of the type (3) with $A_5=3.0 \pm
0.5eV/(s \cdot K^5)$
and $f$=30 MHz, where $f$ is the
SAW frequency (see curve $1$ in Fig.4), and
$A_5=4.0\pm0.6 eV/(s \cdot K^5)$ and $f$=150 MHz
(see curve $2$ in Fig. 4).

\section{THEORY OF HEATING OF TWO-
DIMENSIONAL
ELECTRONS WITH CONTROL OF RELAXATION ON THE
LATTICE BY ELECTRON-ELECTRON COLLISIONS}

To describe the heating of an electron gas by
means of a temperature $T_e$ different from the
lattice temperature $T$, the electron-electron
collisions must occur more often than collisions
with the lattice; i.e., the condition $\tau_{ee}
<< \tau_{\epsilon}$, must be satisfied. Here
$\tau_{ee}$ and $\tau_{\epsilon}$ are,
respectively, the electron energy relaxation time
on phonons and the electron-electron ($ee$)
interaction time.

In a weakly disordered 2DEG in GaAs/AIGaAs
hetero-
structures, momentum is dissipated mainly on the
Coulomb charge of the residual impurity near the
interface. As a result, the relaxation times
satisfy the inequalities

\begin{eqnarray}
\tau_p \ll \tau_{ee} \ll \tau_{\epsilon},
\label{eq:4}
\end{eqnarray}

where $\tau_p$ is the electron momentum
relaxation time.

\subsection{Static regime}

When the inequalities (4) are satisfied, the
nonequilibrium part of the distribution function
has the form

\begin{eqnarray}
f_p=-e {\bf E} \cdot {\bf v} \tau_p
\frac{\partial
f_0(\epsilon_p)}{\partial\epsilon_p},
\label{eq:5}
\end{eqnarray}

where ${\bf E}$ is the electric field, ${\bf v}$
is
the electron velocity, $f_0 (\epsilon_p)$ is the
principal part of the distribution function of
electrons with energy $\epsilon_p=p^2/2m$, where
$p$ and $m$ are, respectively, the electron
momentum and effective mass. Because of the rapid
$ee$ collisions, a Fermi distribution is
established for $f_0(\epsilon_p)$, but the Fermi
level $\epsilon_F$  and the temperature $T_e$
must
be determined from the conservation equations for
the electron density and average energy, while
the
electron-phonon collisions give rise to energy
transfer from the electrons to the lattice.

The results of a calculation of the energy
balance
equation in a 2DEG in the case of electron
scattering by piezoelectric material and
deformation potentials of the acoustic phonons
are
presented in \cite{16,17}. The numerical
coefficients in the relations, taken from
\cite{16} and presented below, refer to a 2DEG on
the (001) surface of GaAs if the following
condition is satisfied:

\begin{eqnarray}
k_F<\pi/d,
\label{eq:6}
\end{eqnarray}

where the electron localization width $d$ in a
quantum well can be estimated for a
heterojunction
by the relation

\begin{eqnarray}
d=[(\frac{3}{4})\frac{a_B^*}{\pi N^*}]^{1/3},
N^*=N_{depl}+\frac{11}{32}n_s.
\label{eq:7}
\end{eqnarray}

Here $N_{depl}$ is the density of the residual
impurity near the heterojunction, and
$a_B^*=\hbar^2\epsilon_sme^2$ is the effective
Bohr radius.

In the case of weak screening, the intensity of
the energy losses due to PA scattering is
determined by the expression \cite{16}

\begin{eqnarray}
Q_{PA}=b_1Q_1(\frac{k_BT}{\hbar k_Fs_t})^3
(\frac{T_e^3}{T^3}-1),
\nonumber \\
Q_1\equiv\frac{2ms^2_t}{\tau_0},
b_1=\frac{\zeta(3)}{2}\frac{13}{16}[1+\frac{9}{13
}
(\frac{s_t}{s})^2],
\label{eq:8}
\end{eqnarray}

where
$1/\tau_0=(e\beta_{14})^2m/2\pi\rho\hbar^2s_t$,
$\beta_{14}$ is the piezoelectric constant,
$\rho$
is the density of the semiconductor (in our case
GaAs, $s$ and $s_t=0.59s$ are, respectively, the
longitudinal and transverse sound speeds in GaAs,
$k_F=(2\pi n_s)^{1/2}$ is the wave number of an
electron with Fermi energy $\epsilon_F$,
$\zeta(x)$ is the Riemann $\zeta(x)$-function,
and
$k_B$ is Boltzmann's constant.

In the case of electron scattering by the
deformation potential of acoustic phonons (DA
scattering) the corresponding expression has the
form

\begin{eqnarray}
Q_{DA}=b_2Q_2(\frac{k_BT}{2ms^2})^2
(\frac{k_BT}{\hbar k_Fs})^3
(\frac{T_e^5}{T^5}-1),
\nonumber \\
Q_2\equiv\frac{2ms^2}{l_0/s},
b_2=12\zeta(5),
\label{eq:9}
\end{eqnarray}

where $l_0 \equiv \pi\hbar^4\rho/2m^3E_1^2$, and $E_1$
is
the deformation
potential.

The relations (8) and (9) hold for small-angle
scattering when

\begin{eqnarray}
k_BT\ll 2\hbar k_Fs \equiv k_B T_{sma},
\label{eq:10}
\end{eqnarray}

and weak screening when

\begin{eqnarray}
k_BT\gg 2\hbar s_t/a_B^*\equiv k_BT_{scr}.
\label{eq:11}
\end{eqnarray}

In the case of strong screening, when an
inequality opposite to the (11) holds,

\begin{eqnarray}
k_BT\ll 2\hbar s_t/a_B^*\equiv k_BT_{scr},
\label{eq:12}
\end{eqnarray}

for PA scattering \cite{16}.

\begin{eqnarray}
Q_{PA}^{scr}=\zeta(5)\frac{3}{4}\frac{59}{64}[1+
\frac{45}{59}(\frac{s_t}{s})^4]
\frac{2ms_t^2}{\tau_0}
\frac{\epsilon_F}{\epsilon_B}
\nonumber \\
\times (\frac{k_BT}{\hbar k_Fs_t})^5
(\frac{T_e^5}{T^5}-1),
\label{eq:13}
\end{eqnarray}

where $\epsilon_B =\hbar^2/2m(a^*_B)^2$ is the Bohr
energy.

\subsection{Heating of electrons by a surface
acoustic wave}

When the relations (4) between the times are
satisfied, the nonequilibrium part of the
distribution function, which depends on the
electron momentum, relaxes rapidly and its
current
part, which is antisymmetric in the momentum, has
the usual form (5) but $E(x,t)=E_o cos(qx-\omega
t)$,
where
$\omega=2\pi f$. As a result, $f_0(\epsilon_p)$
is the Fermi
function
but the chemical potential $\epsilon_F(x,t)$ and
temperature
$T(x,t)$ can be functions of the coordinates and
time. These functions must also be determined
from
the conservation equations for the density and
average energy of the electrons. Slow electron-
phonon ($e-ph$) collisions, which are responsible
for energy transfer from electrons to the
lattice,
appear only in the last equation and they fall
out
of the equation for the density, since the $e-ph$
interaction preserves the total number of
electrons.

The main part of the chemical potential is given
by the normalization condition for the total
electron density, i.e., it is a constant. True,
there are corrections, which are proportional to
the amplitude of the wave, but the nonlinear
contribution from
these corrections, scaled to the main value of
the
chemical potential, is small and can be ignored.
For this reason, we write only the equation for
the change in the average energy

\begin{eqnarray}
\frac{\pi^2}{6}\rho\frac{\partial
(T_e^2)}{\partial t}-\sigma_{xx} E_0^2
\frac{\omega^2}{\omega^2+(q^2D)^2}cos^2(qx-\omega
t) \nonumber \\
+\bar{Q}(T_e)=0,
\label{eq:14}
\end{eqnarray}

where $T_e$ is the electron temperature,
$\rho_0$ is the
two-dimensional density of states, $\sigma_{xx}$
is the
electric conductivity, $D$ is the diffusion
coefficient, and $\bar{Q}(T_e)$ is the energy
transferred
to the lattice. The harmonic variations of the
chemical potential with wave number $q$ and
frequency $\omega$  lead to a variation of the Joule heat
source for the wave and to the appearance in it
of
the cofactor

\begin{eqnarray}
\frac{\omega^2}{\omega^2+(q^2D)^2}.
\nonumber
\end{eqnarray}

Since in the experiment $q^2D \ll \omega$ (see
\cite{7}), the
spatial variation of the Joule heat source can be
disregarded. For this reason, we also disregard
the spatial variation of the temperature but
allow
for a variation of the temperature correction for
the average energy in time. The quantity
$\bar{Q}(T_e)$
depends on the $e-ph$ interaction mechanism. For
PA
scattering $\bar{Q}(T_e)=n_s Q_{PA}(T_e)$, where
$Q_{PA}(T_e)$ are
given by Eq. (8) or (13) and in a simplified form
by the expression (2) or (3);
$n_s=\rho_0\epsilon_F$ is the
total
density of the
two-dimensional electrons.

We shall examine first the condition for weak
heating

\begin{eqnarray}
\Delta T=T_e-T \ll T.
\label{eq:15}
\end{eqnarray}

In this case

\begin{eqnarray}
\frac{\partial \Delta T}{\partial t}+\frac{\Delta
T}{\tau_{\epsilon}}=\frac{3 \sigma_{xx}
E_0^2cos^2(qx-\omega t)}{\pi^2 \rho_0 T}
\label{eq:16}
\end{eqnarray}

where for small-angle PA scattering under strong
screening conditions

\begin{eqnarray}
\frac{1}{\tau_{\epsilon}}=\frac{15}{\pi^2}
\epsilon_F
A_5 T^3,
\label{eq:17}
\end{eqnarray}

and the coefficient $A_5$ is determined by Eq.
written in the form (3). The equation (16) is
easily solved. The temperature correction
nonlinear in the electric field must be
substituted into the expression for the
electrical
conductivity and the latter into the expression
for the damping coefficient $\Gamma$ of the surface
acoustic wave

\begin{eqnarray}
\delta \Gamma=\Gamma (W)- \Gamma
_0=\frac{\partial \Gamma }{\partial \sigma_{xx}}
\frac{\partial \sigma_{xx} }{\partial T} \nonumber \\
\times \frac{3\sigma_{xx}E_0^2\tau_{\epsilon}}{2 \pi^2
\rho_0 T}
(1+\frac{1/2}{1+4\omega^2\tau_{\epsilon}^2}).
\label{eq:18}
\end{eqnarray}

Here $\Gamma _0 \equiv \Gamma(T)$ as $W
\rightarrow 0$ is the absorption in the
linear region at fixed lattice temperature $T$,
and
$\delta \Gamma$ is the nonlinear correction to
$\Gamma (W)$. The SAW
electric field is expressed in terms of the input
power $W$ and the absorption $\Gamma$ is
expressed as
$\sigma_{xx}E_0^2=4\Gamma W$
It follows from the expression (18) that when
$\omega\tau_{\epsilon} \geq 1$ and Eq.
(15) holds, the second
harmonic
in the heating function decreases rapidly as a
result of oscillations in time, and the heating
is
determined by the average power
of the wave. This last assertion is also
valid for the case of strong heating. The
quasistatic balance condition holds in this case:

\begin{eqnarray}
A_5(T_e^5-T^5)=\sigma_{xx}E_0^2/2n_s.
\label{eq:19}
\end{eqnarray}

The temperature $T_e$ found from the relation
(19)
determines the electrical conductivity and the
absorption of the SAW. For strong heating, the
difficulty of solving the nonlinear equation (14)
analytically makes it impossible to obtain simple
formulas for an arbitrary value of the parameter
$\omega\tau_{\epsilon}$.

For $\omega\tau_{\epsilon} \ll 1$, the heating of
the 2DEG is
completely
determined not by the average power but by the
instantaneously varying field of the wave. As a
result, in the case of slight heating, we see an
increase in the degree of heating of the 2DEG
[see
the cofactor in parentheses in the expression
(18)
for $\omega\tau_{\epsilon} \rightarrow 0$]. For
$\omega\tau_{\epsilon} \rightarrow 0$ the
following expression
can be written out, assuming the time derivative
in the relation (16) to be a small term. For the
PA interaction under strong screening conditions

\begin{eqnarray}
T_e(x,t)=[T^5 + \frac{\sigma_{xx} E_0^2 cos^2(qx-\omega t)}{A_5 n_s}]^{1/5}.
\nonumber
\end{eqnarray}

This expression must be substituted into the
temperature dependent part of the electric
conductivity, which in a strong magnetic field is
determined by the expression for the Shubnikov
oscillations

\begin{eqnarray}
\Delta\sigma_{xx}=C \frac{2 \pi^2 T_e(x,t)/ \hbar
\omega_c} {sinh[2 \pi^2 T_e(x,t)/ \hbar
\omega_c]}cos(\frac{2\pi\epsilon_F}{\hbar\omega_c}),
\nonumber
\end{eqnarray}

where $C$ is a slowly varying function of
temperature and magnetic field, and $\omega_c$ is
the
cyclotron frequency. In this case, only the part
of the current corresponding to the first
harmonic
in the 2DEG layer participates in the absorption
of the SAW. The effective temperature appearing
in
the expression for $\Gamma(W)$ is also determined
correspondingly:

\begin{eqnarray}
\frac{T_e}{\sinh(2 \pi^2 T_e/\hbar\omega_c)}  =
 \int_0^{2\pi} \frac{d\varphi}{\pi}(\cos^2\varphi)\nonumber & \\
\times \frac{[T^5+(\sigma_{xx}E_0^2/A_5n_s)\cos^2\varphi]^{1/5}}
{\sinh[(2\pi^2/\hbar\omega_c)[T^5+(\sigma_{xx}
E_0^2/A_5n_s)\cos^2\varphi]^{1/5}]}.
\nonumber
\end{eqnarray}

This expression is quite difficult to use in the
case of strong heating.

\subsection{Determination of the relaxation times}

IV.C.1. {\it Electron-electron interaction time $\tau_{ee}$}.
In the theoretical studies \cite{18,19} it was
shown that
the quasiparticle lifetime in a 2DEG under
conditions of large momentum transfers is
determined by the quantity

\begin{eqnarray}
\frac{\hbar}{\tau_{ee}^{(p)}}=
\frac{\pi^2 T^2}{2\epsilon_{F0}} ln (\frac{\epsilon_{F0}}{T_m}),
T_m=max(T, \hbar/\tau_p),
\label{eq:20}
\end{eqnarray}

where $\epsilon_{F0}$ is the Fermi energy at $T=0$, and $\tau_{ee}^{(p)}$
is
called the "pure" electron-electron ($ee$)
interaction time.

As the temperature is lowered, the so-called
"dirty" or "Nyquist" time $\tau_{ee}^{(N)}$  with small momentum
transfer (in the process of electron diffusion)
$\Delta q \approx 1/L_T$, (\cite{20} and \cite{22}) where
$L_T=(D\hbar/k_BT)^{1/2}$ is
the diffusion length over time $k_BT /\hbar$, often
called
the coherence length, plays an increasingly
larger
role in the $ee$ interaction as the degree of
disordering of the 2DEG increases. The $ee$
collision frequency is determined by the quantity

\begin{eqnarray}
\frac{\hbar}{\tau_{ee}^{(N)}}=
\frac{TR_{\Box}e^2}{h} ln (\frac{h}{2e^2R_{\Box}}),
\label{eq:21}
\end{eqnarray}

where $R_{\Box}=1/\sigma_{xx}$ is the resistance of the film per
unit area.

IV.C.2. {\it Relaxation time
$\bar{\tau}_{\epsilon}$
of the average electron energy}. If the heating
of
the 2DEG is characterized by an electron
temperature $T_e$, then the energy losses $Q$ (per
electron) can be written in the form \cite{10}

\begin{eqnarray}
Q=[\bar{\epsilon}(T_e)- \bar{\epsilon}(T)]/ \bar{\tau}_{\epsilon}
\label{eq:22}
\end{eqnarray}

where $\bar{\epsilon}(T_e)$ and $\bar{\epsilon}(T)$ are the average electron
energy at $T_e$ and $T$, respectively, and $\bar{\tau}_{\epsilon}$
 is the
energy relaxation time. The change in the average
kinetic energy of a two-dimensional electron with
$\epsilon_F \ll k_BT$ is

\begin{eqnarray}
\Delta\epsilon=\bar{\epsilon}(T_e)- \bar{\epsilon}(T)= \frac{\pi^2k_B^2}{6}
\frac{(T_e^2-T^2)}{\epsilon_{F0}}
\mid_{\Delta T \ll T} \nonumber \\
=\frac{\pi^2k_B^2}{3} \frac{T\Delta T}{\epsilon_{F0}}
\label{eq:23}
\end{eqnarray}

The latter equality in Eq. (23) corresponds to
the
condition of weak heating (15). If a dependence
$Q(T_e ,T)$ of the type (2) or (3) can be
represented
in an expansion in $\Delta T/T$ as

\begin{eqnarray}
Q(T,\Delta T)=\gamma A_{\gamma} T^{\gamma -1}\Delta T,
\nonumber
\end{eqnarray}

where $\gamma$ is the exponent of $T_e$ and $T$ in the
expression the following expression (2) or (3),
then we obtain the following expression for $\bar{\tau}_{\epsilon}$:

\begin{eqnarray}
\bar{\tau}_{\epsilon}
%\rule{0.2mm}{5mm}
\mid_{\Delta T \ll T}
=\frac{\pi^2k_B^2}{3\gamma A_{\gamma}\epsilon_{F0} T^{\gamma -2}}.
\label{eq:24}
\end{eqnarray}

For the case (3), i.e., $\gamma=5$, we obtain the
expression (17) for $1/\bar{\tau}_{\epsilon}=1/\tau_{\epsilon}$.

\section{DISCUSSION OF THE EXPERIMENTAL
RESULTS}

Let us examine the condition of applicability of
the heating theories presented in the preceding
section to our results. The typical values of the
residual impurity density $N_{depl}$ in the region of
the 2DEG for our heterostructures is of the order
of $10^{10} cm ^{-2}$. Therefore $N_{depl} \ll n_s$.
For the parameters of
GaAs
$m = 0.07m_0$, permittivity $\epsilon_s=12.8$, and Bohr
radius $a^*_B=97\AA$, we obtain from Eq. (7)

\begin{eqnarray}
d=85\AA, dk_F/\pi \cong 0.3<1
\nonumber
\end{eqnarray}

In other words, the condition (6) is satisfied.

The momentum relaxation time for the experimental
sample was estimated from the Hall mobility
$\tau_p\simeq\mu_Hm/e$, it is $\tau_p=5.1 \cdot10^{-12} s$.

It was shown experimentally in \cite{23} that at
liquid-helium temperatures and low 2DEG
mobilities
the $ee$ interaction with small momentum transfer (21)
predominates in quantum wells at the GaAs/GaAlAs
heterojunction. For our structure, with $R_{\Box}=73 \Omega$,
$\hbar/\tau_{ee}^{(N)} = 1.46 \cdot 10^{-2} T$ and varies in the range

\begin{eqnarray}
\hbar/\tau_{ee}^{(N)}=0.02-0.06 K
\label{eq:25}
\end{eqnarray}
for $T=1.5-4.2K$.

In the expression (20) we employed the value
$T_m=T$,
since $\hbar/\tau_p \simeq 1.5K \leq T$.
In the case $\epsilon_{F0} \simeq 266K$, for
our sample $\hbar/\tau^{(p)}_{ee}$ in the same temperature range is

\begin{eqnarray}
\hbar/\tau_{ee}^{(p)}=0.07-0.4 K
\label{eq:26}
\end{eqnarray}

The sum of the contributions (25) and (26) gives
for the experimental sample

\begin{eqnarray}
1.5 \cdot 10^{-11} s <\tau_{ee}<8.4 \cdot 10^{-11} s
\label{eq:27}
\end{eqnarray}
in the interval $T=1.5-4.2K$.

To estimate the energy relaxation time $\bar{\tau}_{\epsilon}$ (24) it
is necessary to know the coefficient $A_{\gamma}$ in
relations of the type (2) or (3):

\begin{eqnarray}
Q=A_{\gamma}(T_e^{\gamma}-T^{\gamma}).
\nonumber
\end{eqnarray}

A calculation according to Eqs. (8), (9), and
(13)
gives for a 2DEG in our structure [$\beta_{14}=0.12 C/m^2$
(\cite{24}) and the same values of all other
parameters as in \cite{16}] for small-angle
scattering and weak screening, when $ T_{scr} \ll T \ll T_{sma}$ [see
Eqs. (10), (11), (8), and (9)],

\begin{eqnarray}
Q_{PA}=67.5[eV/(s \cdot K^3)](T_e^3-T^3) \nonumber \\
Q_{DA}=13.7[eV/(s \cdot K^5)](T_e^5-T^5),
\label{eq:28}
\end{eqnarray}

and in the case of strong scattering with
$ T\ll T_{scr}  \ll T_{sma}$
[see Eqs. (10), (12), and (13)]

\begin{eqnarray}
Q_{DA}^{scr}= 16.2 [eV/(s \cdot K^5)](T_e^5-T^5),
\label{eq:29}
\end{eqnarray}

As indicated in Ref. 16, PA scattering in the
region of strong screening predominates with
"certainty" over DA scattering.

It should also be noted that for such a sample
with $n_s=6.75\cdot 10^{11} cm^{-2}$ and $\mu_H=1.5\cdot 10^5 cm/(V \cdot s)$
in
dc investigations (i.e., in the static
regime) \cite{11,14} at $T= 1.86 K$ up to $T_e \simeq 4K$ the heating was
described
by a law of the type (2), which is valid for PA
scattering, and under weak screening conditions
the value $A_3=130eV/(s \cdot K^3)$ was found for sample $1$
from \cite{11}, which is higher than the value
indicated for $Q_{PA}$ in Eq. (28) \footnote{ As V.
Karpus has shown \cite{16}, the experimental data
of
\cite{11} in the region $T_e \gg T$ fall well within the
general picture of $Q(T_e,T)$ (see Fig.4 \cite{16}).
It should be noted that the value $\beta_{14}=0.12 C/m^2$ (\cite{24}),
which we used for calculation of
Eqs. (28) and (29), corresponds to $h_{14}= 1.06 \cdot 10^7 V/cm$
(in the notation of \cite{16}). For this
reason, the theoretical value of $A_{\gamma}$ ($\gamma=$3 or 5) in
\cite{11,14}, and \cite{16} for PA scattering (see,
for example, $\alpha \equiv I_3$, for the theoretical curve in
Fig. 3 from \cite{11}) is 1.3 times higher than
the
corresponding values for $Q_{PA}$ presented in the
relations (28) and (29), for similar values of $n_s$ }. However,
irrespective of the
values of $A_{\gamma}$ and $\gamma$ which we used to estimate
$\bar{\tau}_{\epsilon}$,
on
the basis of Eq. (24)-the theoretical values (28) and
(29) or the experimental value $A_3=130eV/(s \cdot K^3)$
$\gamma=3$ - we obtained for the energy relaxation time
estimates in the range
$\bar{\tau}_{\epsilon}=(2-50)10^{-9} s$.

Comparing the values presented above for $\tau_p$ and
$\tau_{ee}$ (27) and the range of values for $\bar{\tau}_{\epsilon}$,
we see
that the relations (4) are satisfied.The concept
of an electron temperature $T_e$ could therefore be
introduced and the heating theories presented in
Sec.IV could be used.

Let us examine the estimates of the critical
temperatures $T_{sma}$ (10) and $T_{scr}$ (12) at which
the
energy relaxation mechanisms change in the case
of
the $e-ph$ interaction. We
determined   these   temperatures   using   the
value $s_t=3.03 \cdot 10^5 cm/s$ (see \cite{16}) and the
value given above for $a^*_B$ . The results are

\begin{eqnarray}
T_{sma}=9.5K \mbox{ and } T_{scr}=4.6K
\label{eq:30}
\end{eqnarray}

Since the phonon temperature in our experiments $T=1.55K$, we have

\begin{eqnarray}
T < T_{scr} < T_{sma}
\label{eq:31}
\end{eqnarray}

Therefore, the inequalities (10) and (12) are
satisfied in our experiment, though not as
strongly, especially the inequality (12), as
assumed in the theory of \cite{16} for application
of the expression (13).

Finally, observation of a law of the type (3)
with
$\gamma=5$ (see Sec.III and Fig. 4) and the ratio (31)
of
the temperatures presented above allows us to
assert that in the case of heating of two-
dimensional electrons by the electric field of a
SAW ($f$=30 and 150 MHz) the electron energy
relaxation is determined by PA scattering with
strong screening (13), which for the parameters
employed by us gives the theoretical relation
(29).

At the same time, as noted above, in the
investigation in the static regime [with phonon
temperature $T \simeq 1.86K$ \cite{11,12,13,14,15}, i.e. the
inequalities (31) hold], the law (2) with $\gamma=3$ was
observed, indicating that PA scattering dominates
in the electron energy relaxation mechanisms in
the case of weak screening (8). Besides the
indicated discrepancy between the results of
investigations of the heating of a 2DEG in high-
frequency (rf) and dc electric fields, it should
be noted that there is also a discrepancy in the
experimental values $A_5 \simeq 3eV/(s \cdot K^5)$ at $f$=30MHz and
values $A_5 \simeq 4 eV/(s \cdot K^5)$ at $f$=150 MHz (see Sec.III). In
addition, these values are not greater than (as
the experimental value of $A_{\gamma}$ is the static
regime)
but   less   than   the   theoretical   value
values $A_5 \simeq 3eV/(s \cdot K^5)$ - Eq. (29), calculated according
to
the theory of \cite{16}.

Since the calculations in \cite{16} were performed
for a constant electric field, they obviously
cannot explain the above-noted discrepancies,
especially the difference in the functions $Q(T_e)$
at different frequencies. Apparently, the
difference is due to the different values of
$\omega\tau_{\epsilon}$
with respect to 1. Taking into consideration the
approximate nature of the computed parameters and
the uncertainty in the input power in our
measurements, we took as the value of the energy
relaxation time $\tau_{\epsilon}$ estimated from the theoretical
value $A_5 \simeq 16.2 eV/(s \cdot K^5)$ with $\gamma=5$ (29), which
gives in the case
of a calculation based on Eq. (17) or (24)
$\tau_{\epsilon}\simeq 3.3 \cdot 10^{-9} s$.
At frequency $f=30$ MHz $\omega\tau_{\epsilon} \simeq 0.6<1$ and
at $f=150$MHz $\omega\tau_{\epsilon} \simeq 3 > 1$, which leads to a different
heating for the same energy losses. In this
connection, we attempted to study this question
theoretically (see Sec.IV.B) and to compare the
results obtained with experiment. As a result, we
can demonstrate the validity of Eq. (18),
obtained
under the assumption of weak heating, $\Delta T \ll T$. We
present in the inset
in Fig. 4 the experimental values of the
difference $\delta\Gamma=\Gamma(W)-\Gamma_0$ as a function of $Q$ for
two frequencies, 30 and 150 MHz, in a field $H=
15.5$ kOe. We see from the figure that in
accordance with Eq. (18), these dependences are
linear and for the same energy
losses $Q$ the quantity $\delta\Gamma_1$ ($f$=30 MHz) is
greater than $\delta\Gamma_2$ ($f_2$=150 MHz),
the ratio $\delta\Gamma_1/\delta\Gamma_2$  is
equal, to within $10\%$, to the theoretical value

\begin{eqnarray}
(1+\frac{1/2}{1+4\omega_1^2\tau_{\epsilon}^2})/
(1+\frac{1/2}{1+4\omega_2^2\tau_{\epsilon}^2})
\nonumber
\end{eqnarray}

with $\omega_{1,2}=2\pi f_{1,2}$. A similar result was also
obtained for $\delta\Gamma_1/\delta\Gamma_2$
in the magnetic field $H=14.1$ kOe.
Therefore, experiment confirms the
theoretical conclusion that for $\Delta T \ll T$ the energy
losses depend on  $\omega\tau_{\epsilon}$.

It should be noted that in determining $Q$ at $f=$150
MHz it was assumed that $\delta[\Gamma (W)/\Gamma_M]$ is
frequency-
independent (see Sec.III), which is at variance
with the result presented above. However, $Q$ and
$\delta\Gamma$
are so small at the onset of the nonlinear
effects
that
their differences at different frequencies fall
within the limits of error of our measurements.

As one can see from the theory (see Sec. IV.B), an
analytical expression could not be obtained in
the
case of strong heating of a 2DEG in an rf
electric
field of a SAW, but it can be assumed that, by
analogy with the case of weak heating, the
difference in the coefficients $A_5$ remains also in
the case of heating up to $T \simeq 4$.

A more accurate numerical development of the
theory of heating of a 2DEG for arbitrary values
of  $\omega\tau_{\epsilon}$ from  $\omega =0$ up to
$\omega\tau_{\epsilon} \gg 1$, including in the
transitional regions $T \simeq T_{scr}$ and $T \simeq T_{sma}$, could
explain the discrepancy in the experimental
results obtained in constant and rf electric
fields with the same direction of the
inequalities
(31).

\section{CONCLUSIONS}

In our study we observed heating of a 2DEG by a
rf
electric field generated by a surface acoustic
wave (SAW). The heating could be described by an
electron temperature $T_e$, exceeding the lattice
temperature $T$.

It was shown that the experimental dependences of
the energy losses $Q$ on $T_e$ at different SAW
frequencies depend on the value of  $\omega\tau_{\epsilon}$  with
respect to $1$, where $\tau_{\epsilon}$  is the energy relaxation
time of two-dimensional electrons. Theoretical
calculations of the heating of a two-dimensional
electron gas by the electric field of a SAW were
presented for the case of warm electrons ($\Delta T \ll T$).
The results showed that for the same energy
losses
$Q$ the degree of heating (i.e. the ratio $T_e/T$)
with  $\omega\tau_{\epsilon} > 1$($f$=150 MHz) is less than
with  $\omega\tau_{\epsilon} < 1$
($f$=30 MHz). Experimental results confirming this
calculation were presented.

It was shown that the electron energy relaxation
time $ \tau_{\epsilon}$ - is determined by energy dissipation in
the
piezoelectric potential of the acoustic phonons
under conditions of strong screening for the SAW
frequencies employed in the experiment.

This work was supported by the Russian Fund for
Fundamental Research (Grants 95-02-04066a and 95-
02-04042a) and by the Fund of International
Associations (Grant INTAS-1403-93-ext).

\end{multicols}

%\newpage
%\narrowtext
\begin{figure}
\centerline{\psfig{figure=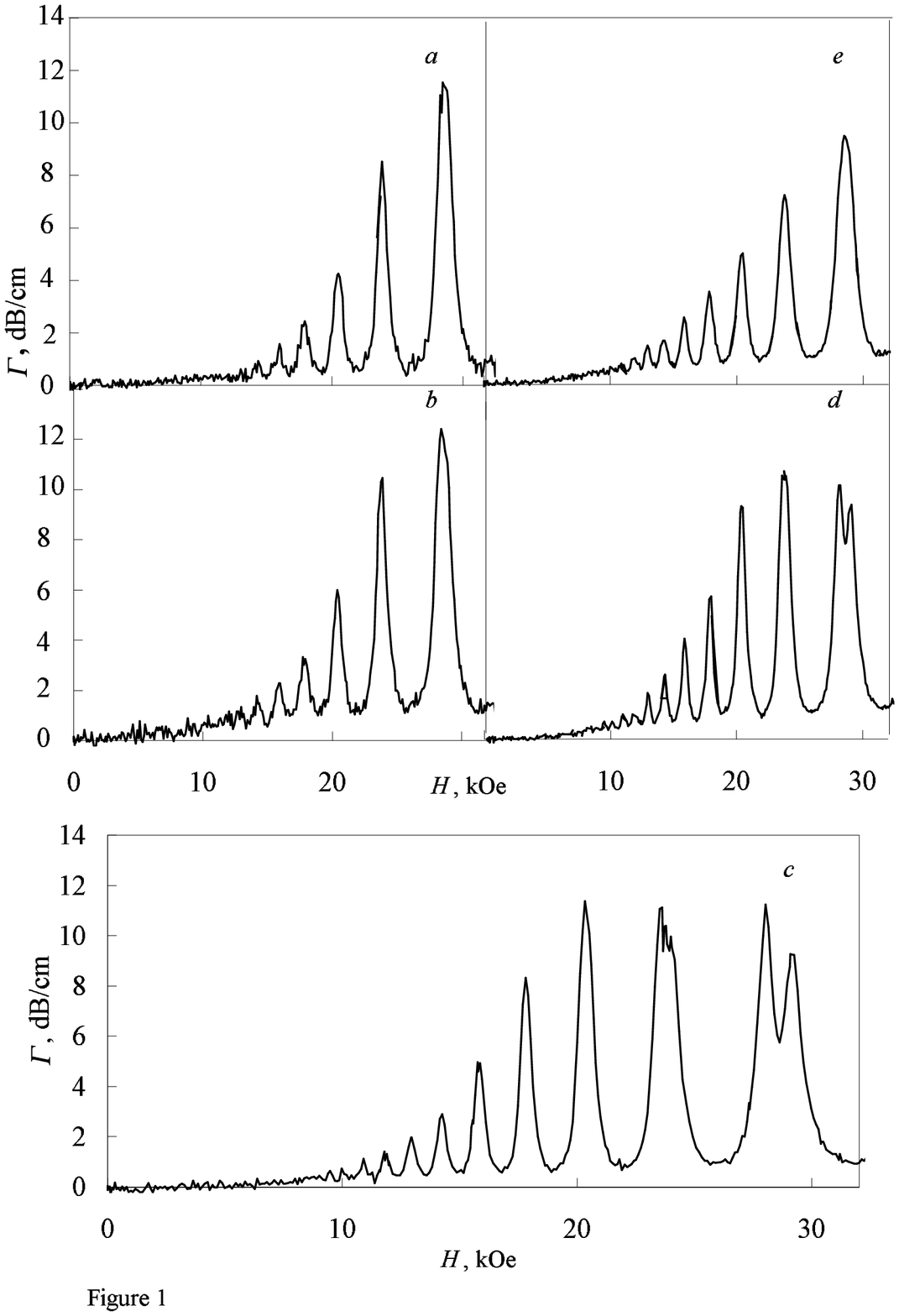,width=9cm}}
\vspace{-0.5cm}
\caption{Absorption coefficient $\Gamma$ versus
magnetic field $H$ at frequency $f$=30 MHz at
temperatures T,K: a-4.2, b-3.8, c-e-1.5 and wave
power at the oscillator output $P$, W: a-c-$10^{-
5}$, d-$10^{-4}$ e-$10^{-3}$.}
\end{figure}

\begin{figure}
%\narrowtext
\centerline{\psfig{figure=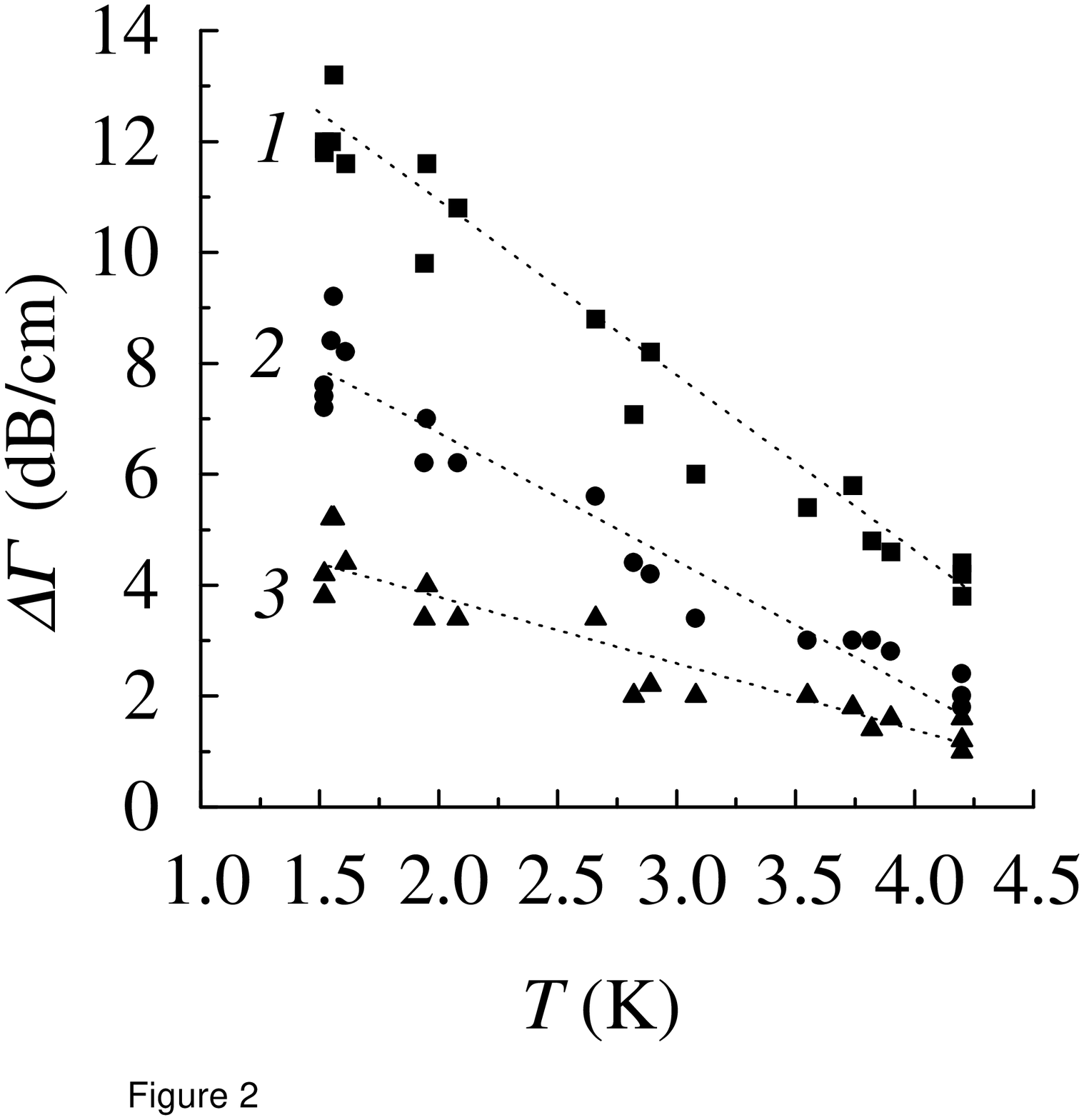,width=9cm}}
\vspace{-0.5cm}
\caption{
$\Delta\Gamma=\Gamma_{max}-\Gamma_{min}$
versus temperature $T$ in the linear regime at a
frequency of 150 MHz in a magnetic field $H$,
kOe: $1$-17.5, $2$-15.5, $3$-14.1.}
\end{figure}

\begin{figure}
%\narrowtext
\centerline{\psfig{figure=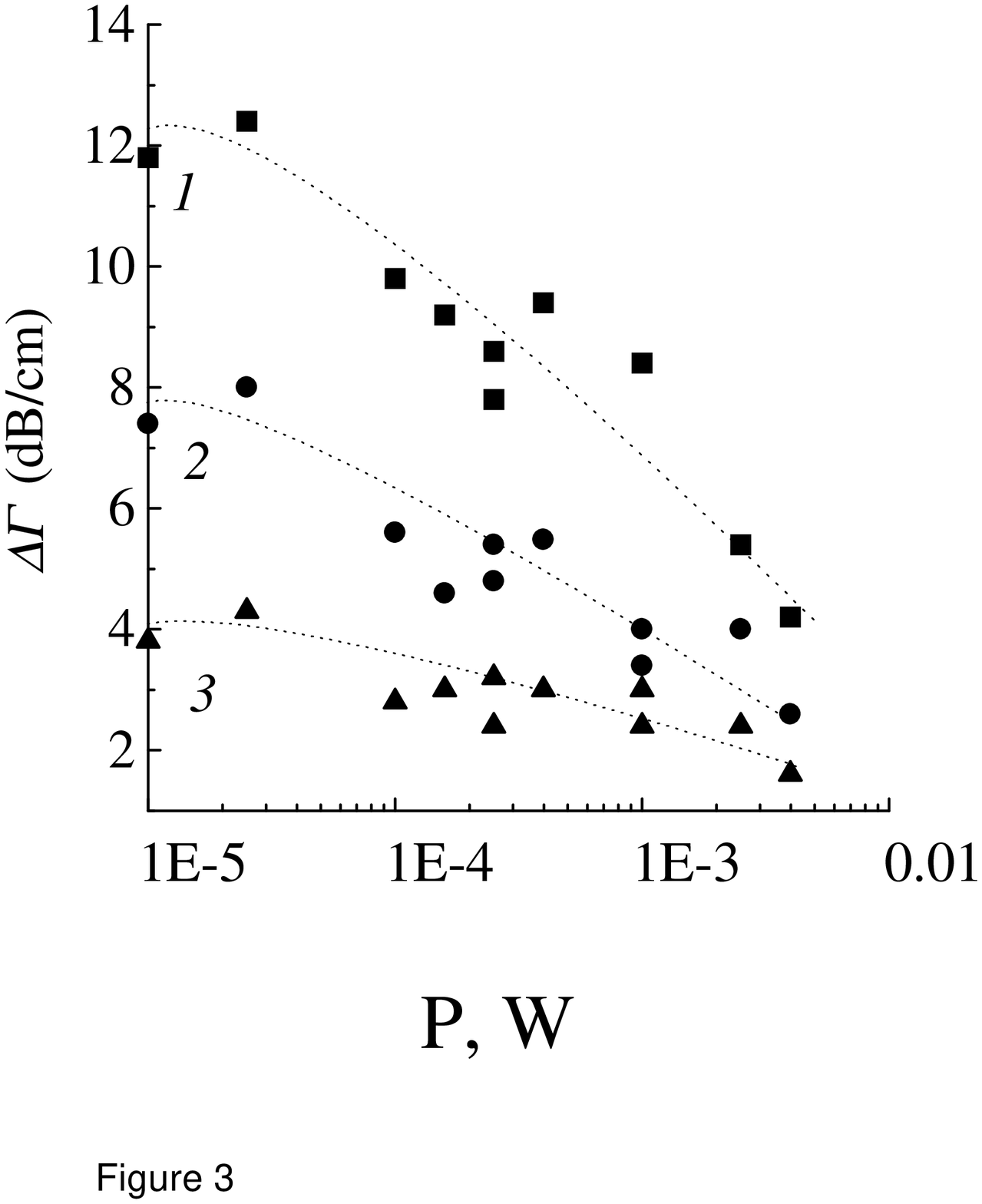,width=9cm}}
\vspace{-0.5cm}
\caption{
$\Delta\Gamma=\Gamma_{max}-\Gamma_{min}$
versus the power $P$ at the oscillator output in
the linear regime at a frequency of 150 MHz in a
magnetic field $H$, kOe: 1-17.5,2-15.5, 3-14.1.}
\end{figure}

\begin{figure}
%\narrowtext
\centerline{\psfig{figure=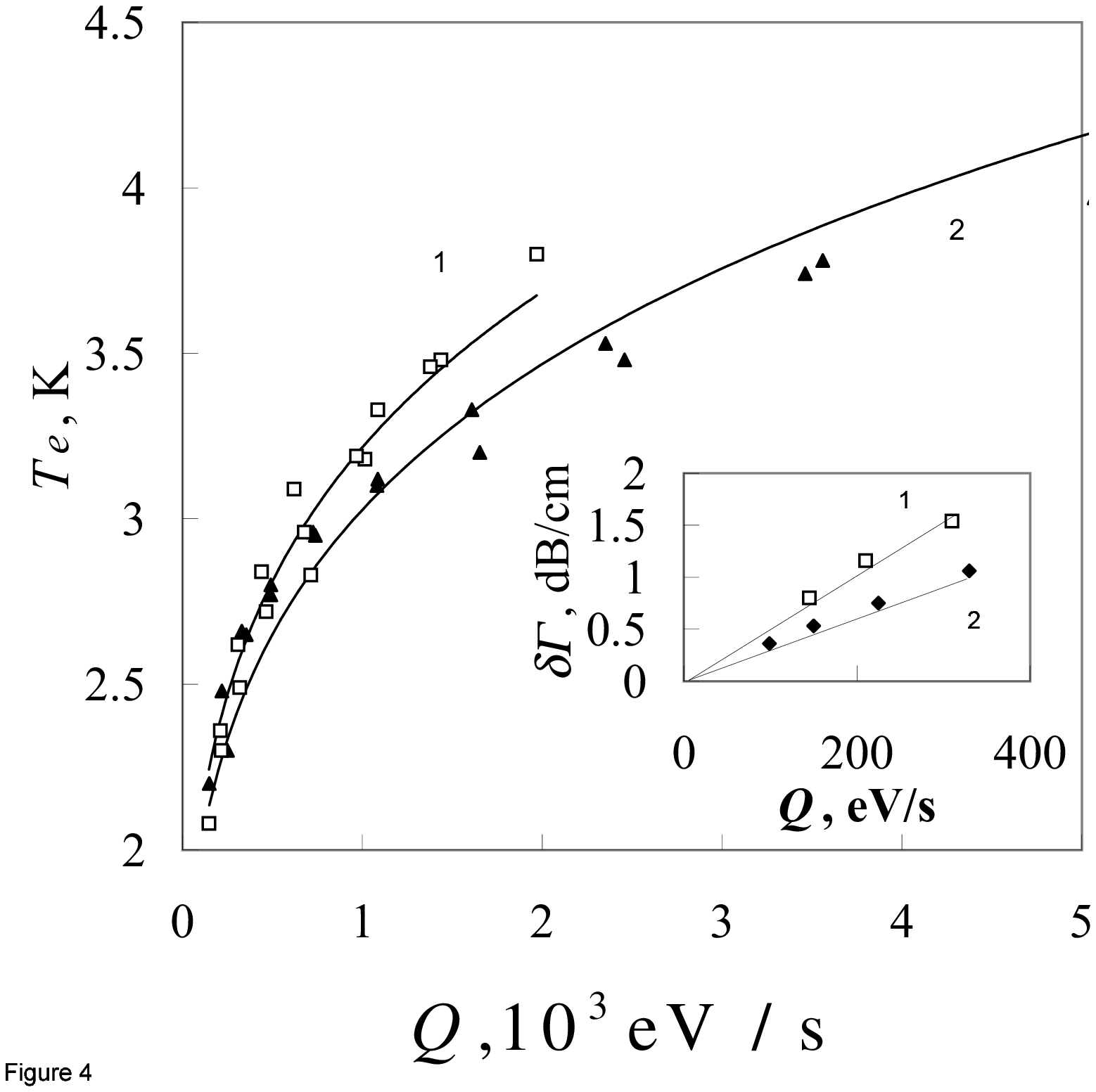,width=9cm}}
\vspace{-0.5cm}
\caption{
Electron temperature $T_e$ versus the
energy losses $Q$ at SAW frequencies $f$, MHz:
$1$-150 and $2$-30. Inset:
$\delta\Gamma=\Gamma(T_e)-\Gamma_0(T)$ versus $Q$
at SAW frequencies $f$ MHz: $1$-150, $2$-30.}
\end{figure}

%\end{multicols}
\end{document}